\documentclass[aps, pra,reprint,showpacs,showkeys]{revtex4-1}
\usepackage{amsmath, amssymb}    % need for subequations
\usepackage{graphicx}   % need for figures
\graphicspath{ {./img/} }
\usepackage{verbatim}   % useful for program listings
\usepackage{color}      % use if color is used in text
\usepackage{subfigure}  % use for side-by-side figures
\usepackage{hyperref}   % use for hypertext links, including those to external documents and URLs
\usepackage{pstricks}
\usepackage{psfrag}
\usepackage[percent]{overpic}

\newcommand{\W}{\mathcal{W}}

\newcommand{\ket}[1]{\ensuremath{|#1\rangle}}

\DeclareMathOperator{\tr}{Tr}

\definecolor{darkgreen}{RGB}{0,100,0}
\begin{document}
\title{Elastic and inelastic transmission in guided atom lasers: a truncated Wigner approach}
\author{Julien Dujardin}
\email{julien.dujardin@ulg.ac.be}
\author{Arturo Arg\"{u}elles}
\author{Peter Schlagheck}
\affiliation{D\'{e}partement de Physique, University of Liege, 4000 Li\`ege, Belgium}

%\date{}
\begin{abstract}
We study the transport properties of an ultracold gas of Bose-Einstein condensate that is coupled from a magnetic trap into a one--dimensional waveguide. Our theoretical approach to tackle this problem is based on the truncated Wigner method for which we assume the system to consist of two semi-infinite non-interacting leads and a finite interacting scattering region with two constrictions modelling an atomic quantum dot. The transmission is computed in the steady-state regime and we find a good agreement between truncated Wigner and Matrix-Product State calculations. We also identify clear signatures of inelastic resonant scattering by analyzing the distribution of energy in the transmitted atomic matter wave beam.
\end{abstract}
\maketitle

\section{Introduction}
The progress of the last two decades in the field of ultracold atoms has opened the possibility of investigating mesoscopic transport properties of interacting matter waves. A very important step in this context is the realization of atom lasers \cite{Bloch1999PRL,Cennini2003PRL,Hagley1999S,Mewes1997PRL} permitting to create a beam of atoms by coherently outcoupling a trapped Bose--Einstein condensate (BEC) into an optical waveguide at a well-defined energy and flux \cite{Guerin2006PRL,Couvert2008EEL,Riou2008PRA,Gattobigio2009PRA,Debs2010PRA,KleineBuening2010APB,Gattobigio2011PRL}. This research is particularly interesting in view of the perspective to realize bosonic atomtronic devices \cite{Micheli2004PRL,Daley2005PRA,Seaman2007PRA,Pepino2009PRL} and to study analogies with their fermionic counterpart \cite{Brantut2012S,Bruderer2012PRA,Kristinsdottir2013PRL}. A typical configuration of such a device is an atomic quantum dot that features resonant transport \cite{Carusotto2001PRA} and atom blockade \cite{Schlagheck2010NJoP,Kristinsdottir2013PRL}. These features can be used as building blocks for atomic transistors \cite{Micheli2004PRL,Seaman2007PRA,Pepino2009PRL}. 

A theoretical modeling of such scattering processes within guided atom lasers faces the challenge of dealing with interactions between atoms. A full many-body treatment of such an open system is very complex and impossible to solve exactly in practice. During last years, these scattering processes have been studied in the mean--field approximation described by a nonlinear Gross--Pitaevskii (GP) equation \cite{Paul2005PRL,Paul2007PRA,Ernst2010PRA}. While this description gives satisfactory results for a weak nonlinearity, the question of validity arises rapidly in the case of strong nonlinear dynamics \cite{Ernst2010PRA} where dynamical instabilities occur. It has also been pointed out that in the presence of disordered potentials, even a weak atom-atom interaction strength can lead to inelastic scattering processes \cite{Geiger2012PRL,Geiger2013NJoP}, which can not be accounted for in the framework of the mean--field GP approximation.

The main focus of this work is to study such inelastic scattering processes in an atom laser context. We employ the truncated Wigner method (tW) \cite{Wigner1931,Wigner1932PR,Moyal1949PCPS} for this purpose. The latter amounts to sampling the initial quantum state by classical fields and to propagating them according to a slightly modified GP equation. This method has been used to study the reflection of a BEC on abrupt potential barriers at zero temperature \cite{Scott2006PRA} and at finite temperature \cite{Scott2007LP}. It can also be used to study the dynamics of a trapped BEC \cite{Isella2006PRA} when an optical lattice is adiabatically superimposed to the trapping potential. It has also been used to study the many-body Landau-Zener effect \cite{Altland2009PRA} as well as far from equilibrium dynamics (in particular non-thermal fixed-points) of many-body systems \cite{Schmidt2012NJoP}. Finally, the tW method can also take into account a continuous measurement process \cite{Lee2014PRA} during the evolution of the system.

We specifically apply the tW method to study the transmission of a one-dimensional guided atom laser beam across a double barrier potential forming an atomic quantum dot as described in Sec.~\ref{sec:scattconf}. In this scenario, we are particularly interested in resonant transport. To this end, we suppose a finite extent of the interacting region and discretize the one-dimensional space according to a finite-difference scheme. We generalize, in Sec.~\ref{sec:tWBH}, the tW method to open systems using Smooth Exterior Complex Scaling \cite{Balslev1971CMP,Simon1973AM,Simon1979PLA,Kalita2011JCP,Dujardin2014APB}. We then study numerically, in Sec.~\ref{sec:TransQD}, the transmission properties through the quantum dot model described in Sec.~\ref{sec:scattconf}. The obtained results are then confronted, in Sec.~\ref{subsec:transmQD}, to the predictions provided by the mean--field approximation and to Matrix-Product State (MPS) calculations. In Secs.~\ref{sec:TransQD}~{B--D}, we analyze the energy distribution of the transmitted beam using the tW method and develop a Bogoliubov approach to understand the physical origin of the inelastic peaks that appear in the energy distribution. 

\section{Scattering configuration} 
\label{sec:scattconf}
\begin{figure}[hbtp]
  \centering
  \begin{psfrags}
    \psfrag{(a)}{\textbf{(a)}}
    \psfrag{(b)}{\textbf{(b)}}    
    \psfrag{waveguide}{waveguide}
    \psfrag{BEC}{\textcolor{darkgreen}{BEC}}    
    \includegraphics[width=0.9\linewidth]{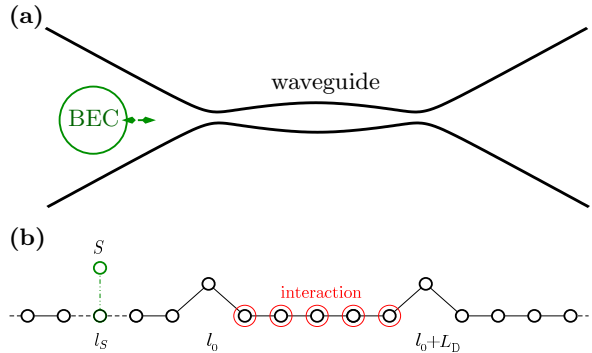}
  \end{psfrags}
  \caption{(color online) (a) A trapped BEC depicted by a (green) circle is loaded into a waveguide with two constrictions modeling an atomic quantum dot. The bold (black) lines represent the isopotentials of the waveguide. (b) One dimensional infinite Bose-Hubbard (BH) chain for the quantum dot model (see Eqs.~\ref{eq:qdot}). The condensate is prepared within a trap represented by the green circle ($S$) and coupled to the infinite BH chain (dashed green line). The big (red) circles represent a non-vanishing on-site atom-atom interaction. The two displaced sites enclosing the interaction region represent two sites where the on-site potential is nonzero.}
\label{fig:qdot_geom}
\end{figure}

We consider a guided atom laser experiment such as the one represented in Ref.~\cite{Guerin2006PRL}, where a magnetically trapped BEC plays the role of a coherent source of atoms. In this particular experiment, the atoms are out-coupled by a rf-knife rendering the final state insensitive to the magnetic field, but sensitive to the optical potential formed by  an elongated far off-resonance optical beam constituting an atomic waveguide. Ideally, the propagation of the atoms at well defined energy is quasi one--dimensional (1D) along the waveguide. It is then possible to engineer an atomic quantum dot geometry by focusing two far--detuned laser beams perpendicular to the waveguide. In this paper, we specifically consider a waveguide configuration in which spatial inhomogeneities and atom-atom interactions are non-vanishing only in a finite region of space. Such a system is represented in Fig.~\ref{fig:qdot_geom}(a). 

In order to properly implement the tW method, we discretize the 1D space by a series of points or sites separated by a constant distance $\Delta$, thereby forming a spatial grid. The wavefunction is then defined on these points. The sites are labeled with an index $l\in\mathbb{Z}$. One additional site $S$ is introduced in order to represent the source of atoms. This additional site $S$ is connected to the waveguide at site $l_S$ as illustrated in Fig.~\ref{fig:qdot_geom}(b).

We treat the spatial derivatives with a finite-difference approximation. The Hamiltonian is then given by 
\begin{eqnarray}
\label{eq:BHham}
	\hat{\mathcal{H}} &= \displaystyle\sum_{l=-\infty}^{+\infty}& \Bigg[ 
-J(\hat{a}^\dagger_{l+1}\hat{a}_{l} + \hat{a}^\dagger_{l}\hat{a}_{l+
1}) \nonumber \\
				&&+ \frac{g_l}{2} \hat{n}_l(\hat{n}_l-1
) + V_l\hat{n}_l \Bigg] \nonumber \\ 
				&&+ \kappa^*(t)\hat{b}^\dagger\hat{a}_{l_S} +
 \kappa(t)\hat{a}^\dagger_{l_S}\hat{b} + \mu \hat{b}^\dagger \hat{b}.
\end{eqnarray}
Here $\hat{b}$ and $\hat{b}^\dagger$ is the annihilation and creation operator of the reservoir, respectively, $\mu$ is its chemical potential defined relative to the center of the band, and $\hat{a}_l$ and $\hat{a}_l^\dagger$ are the annihilation and creation operators, respectively, on the site $l$ of the chain. The hopping strength to the nearest neighbors is given by $J=\hbar^2/2m\Delta^2$, the on-site interaction strength is $g_l$, and the on-site potential is $V_l$. The coupling strength $\kappa(t)$ is related to the out-coupling process of atoms from the reservoir and can be controlled in a time-dependent manner (\textit{e.g.} through the variation of the intensity of a radio-frequency field in the case of Refs.~\cite{Guerin2006PRL,Riou2008PRA}). We suppose that the source is adiabatically switched on from zero to a maximal value of $\kappa$, \textit{i.e.}
\begin{equation}
	\lim_{t\to\infty} \kappa(t) = \kappa.
\end{equation}

This Hamiltonian is similar to a Bose-Hubbard (BH) system describing an optical lattice in which only the lowest band in the Brillouin zone is considered. For the case where the on-site potential and interaction strength vanish, the dispersion relation is identical to the one of the free lattice and is given by 
\begin{equation}
\label{eq:DispRel}
	E(k) = -2J \cos(k),
\end{equation}
with a wavenumber $k/\Delta$. In the limit $k\ll1$, we have $E(k) = -2J + J k^2$ which, apart from a constant shift, corresponds to the dispersion relation of a free atom. 

The scattering configuration of an atomic quantum dot is modeled by two sites with non-zero on-site potential. Between these two sites, we allow atoms to interact as depicted in Fig.~\ref{fig:qdot_geom}(b). This can be justified if, for instance, the waist of the elongated optical beam is particularly narrow at the position where the quantum dot is located. The perpendicular confinement is then rather strong and it is likely that collisions occur between atoms. Formally, this model can be encoded as
\begin{subequations}
\label{eq:qdot}
 \begin{align}
  V_l &= V(\delta_{l,l_0} + \delta_{l,l_0+L_{\mathrm{D}}}), \label{eq:qdotV} \\
  g_l &= g \sum_{j=1}^{L_{\mathrm{D}}-1}\delta_{l,l_0+j}, \label{eq:qdotg}
 \end{align}
\end{subequations}
where $l_0\in\mathbb{Z}$ is arbitrary and $L_{\mathrm{D}}$ is the length of the quantum dot. We set $L_{\mathrm{D}}=6$ in the rest of the paper.

\section{Truncated Wigner Method for open BH systems}
\label{sec:tWBH}

Phase-space methods were introduced by Wigner \cite{Wigner1931,Wigner1932PR} and Moyal \cite{Moyal1949PCPS} and their development started in the 60's with successful applications in quantum optics by Glauber \cite{Glauber1963PR} and Sudarshan \cite{Sudarshan1963PRL}. These methods allow to go beyond the mean--field GP description by, essentially, sampling the initial quantum state by classical fields. The prescription to sample the initial state and the equation of motion are not unique. In this paper we choose the truncated Wigner method (tW). The evolution of the system is then given by a classical equation of motion similar to the GP equation. In particular, the tW method maps the density matrix of the system onto a quasi--distribution function fulfilling a Fokker--Planck equation. It is then possible to replace this equation with a system of Langevin equations that can be numerically solved by a Monte-Carlo method. This section is devoted to generalize the tW method to open systems.

\subsection{Truncated Wigner Method for BH systems}
\label{subsec:tWinfBH}
Let us consider a general Bose-Hubbard (BH) system with on-site two-body interaction. Denoting by $\mathcal{A}=\{S,0,\pm1,\pm2,\cdots\}$ the ensemble of sites of the BH system, the many-body Hamiltonian of the system can be written as
\begin{equation}
\label{eq:MBHBH}
\mathcal{\hat{H}} = \sum_{\alpha\in \mathcal{A} } \Bigg[ \sum_{\alpha'\in \mathcal{A} }h_{\alpha\alpha'}\hat{a}^\dagger_{\alpha}\hat{a}_{\alpha'} + \frac{g_\alpha}{2} \hat{n}_\alpha(\hat{n}_\alpha-1)\Bigg],
\end{equation}
where we defined  by $\hat{a}_\alpha$ and $\hat{a}_\alpha^\dagger$ the annihilation and creation operators, respectively, on the site $\alpha$ of the chain, and by $\hat{n}_\alpha = \hat{a}^\dagger_\alpha\hat{a}_{\alpha}$ the corresponding number operator. The matrix elements $h_{\alpha\alpha'}$ represent on-site energies as well as possible hoppings between the sites. We impose $h_{\alpha\alpha'}=h_{\alpha'\alpha}^*$ to ensure that the Hamiltonian remains hermitian. This general form makes our description also valid for more involved connections between different sites of the grid, such as small-world networks.

The general idea of the Wigner approach is to map the evolution of the density matrix prescribed by the von Neumann equation
\begin{equation}
i\hbar\frac{\partial\hat{\rho}(t)}{\partial t} = [\mathcal{\hat{H}}, \hat{\rho}(t)]
\end{equation}
to the evolution of the Wigner function $\mathcal{W}~\equiv~\mathcal{W}(\{\psi_\alpha,\psi_\alpha^*\},t)$ that is defined in the phase space spanned by the classical amplitudes $\psi_\alpha$ associated with each site $\alpha$. The Wigner function represents a quantum quasi-probability distribution and is defined as 
\begin{eqnarray}
\mathcal{W}(\{\psi_\alpha,\psi_\alpha^*\},t) &=& \prod_{\alpha \in \mathcal{A}}\frac{1}{\pi^2} \iint d\lambda_\alpha d\lambda^*_\alpha e^{-\lambda_\alpha\psi^*_\alpha + \lambda_\alpha^*\psi_\alpha} \nonumber \\  && \times\,\,\, \chi_{\mathcal{W}}(\{\lambda_\alpha,\lambda_\alpha^*\},t),
\end{eqnarray}
which is the Fourier transform of the characteristic function $\chi_{\mathcal{W}}$
\begin{equation}
\chi_{\mathcal{W}}(\{\lambda_\alpha,\lambda_\alpha^*\},t) = \tr\left[\hat{\rho}(t) \prod_{\alpha \in \mathcal{A}} e^{\lambda_\alpha\hat{a}_\alpha^\dagger - \lambda_\alpha^*\hat{a}_\alpha} \right].
\end{equation}
The classical amplitudes $\psi_\alpha$ and $\psi_\alpha^*$ are complex canonical variables representing coherent states in phase space. The evolution of the Wigner function is then given by
\begin{eqnarray}
\label{eq:Wignerevol}
i\hbar\frac{\partial \W}{\partial t}  &=& \sum_{\alpha\in\mathcal{A}} \Bigg[ -\sum_{\alpha'\in\mathcal{A}} \left(h_{\alpha\alpha'}\frac{\partial}{\partial \psi_\alpha}\psi_{\alpha'} - h_{\alpha'\alpha}^* \frac{\partial}{\partial \psi_\alpha^*}\psi_{\alpha'}^* \right)  \nonumber\\ 
				    &-& g_\alpha \left(\frac{\partial}{\partial\psi_\alpha}\psi_\alpha - \frac{\partial}{\partial\psi_\alpha^*}\psi_\alpha^*\right)(|\psi_\alpha|^2-1) \nonumber   \\
				    &+&  \frac{g_\alpha}{4} \left(\frac{\partial^2}{\partial \psi_\alpha^2}\frac{\partial}{\partial \psi^*_\alpha}\psi_\alpha - \frac{\partial}{\partial \psi_\alpha}\frac{\partial^2}{\partial \psi^{*^2}_\alpha}\psi_\alpha^*\right) \Bigg ] \W.	
\end{eqnarray}
Numerical integration of this equation is practically impossible since the dimension of the phase space is very large. 

The so--called \emph{truncated Wigner approximation} consists in neglecting the third order derivatives in Eq.~\eqref{eq:Wignerevol}. The resulting equation is commonly called the \emph{truncated Wigner equation} and corresponds to a Fokker--Planck equation with only a drift term. It can be shown \cite{Gardiner2004} that this approximation is valid if there is locally a large number of atoms in the waveguide. The evolution of the Wigner function can be mapped to a set of coupled Langevin equations where the canonical variables $\psi_\alpha\equiv\psi_\alpha(t)$ and $\psi^*_\alpha\equiv\psi^*_\alpha(t)$ are now time-dependent. They satisfy
\begin{equation}
\label{eq:CanonicalEvolution}
i\hbar\frac{\partial \psi_\alpha}{\partial t} = \sum_{\alpha'\in\mathcal{A}} h_{\alpha\alpha'}\psi_{\alpha'} + g_\alpha(|\psi_\alpha|^2 -1)  \psi_\alpha.
 \end{equation}
The mapping gives another set of equations for the evolution of $\psi_\alpha^*$ which correspond to the complex conjugate of Eq.~\eqref{eq:CanonicalEvolution}.

For the specific case of our guided atom-laser configuration, we can now write the final set of equations of motion for the sites representing the waveguide and the site corresponding to the source as
\begin{subequations}
\label{eq:StochAll}
 \begin{align}
i\hbar\frac{\partial \psi_l}{\partial t} &= (V_l-\mu)\psi_l -J\left(\psi_{l+1} + \psi_{l-1}\right)  \nonumber\\
							 &  + g_l(|\psi_l|^2 -1) \psi_l + \kappa(t)\psi_S\delta_{l,l_S}, \\
\label{eq:StochSource}
i\hbar\frac{\partial \psi_S}{\partial t} &=  \kappa^*(t)\psi_{l_S}.
 \end{align}
\end{subequations}
It is nearly identical to a discrete GP equation except for a slightly different interaction term.
 
\subsection{The initial state}
\label{subsec:tWinit}
The initial Wigner function $\mathcal{W}(\{\psi_\alpha,\psi_\alpha^*\},t_0)$ represents the initial quantum state of the system and has to be sampled by the classical fields $\psi_\alpha$. The latter can, for instance, represent coherent, thermal, squeezed or Fock states \cite{Olsen2009OC} and its time evolution is governed by classical trajectories evolving according to Eqs.~\eqref{eq:StochAll}. We consider that initially, at $t=t_0$, the waveguide is empty and the ground state of the reservoir trap is macroscopically populated with a large number $N$ of atoms at zero temperature. The Wigner function can then be written as 
\begin{equation}
	\W = \mathcal{W}_G(\{\psi_l,\psi_l^*\},t_0) \, \times \, \mathcal{W}_S(\psi_S,\psi_S^*,t_0),
\end{equation}
at time $t=t_0$, where $\mathcal{W}_G(\{\psi_l,\psi_l^*\},t_0)$ and $\mathcal{W}_G(\{\psi_l,\psi_l^*\},t_0)$  correspond to the Wigner function of the source of atoms and the waveguide, respectively.

Since the waveguide is initially empty, the corresponding Wigner function has the form \cite{Sinatra2002JPBAMOP}
\begin{equation}
\mathcal{W}_G(\{\psi_l,\psi_l^*\},t_0) = \prod_{l \in \mathbb{Z}} \left(\frac{2}{\pi}\right) \exp(-2|\psi_l|^2).
\end{equation}
We can therefore sample the initial state with complex Gaussian random variables. More precisely, the initial values of the amplitudes $\psi_l$ can be written as
\begin{equation}
\label{eq:InitCond}
\psi_l(t=t_0) = \frac{1}{2}\left(A_l + i B_l\right),
\end{equation}
where $A_l$ and $B_l$ are real, independent Gaussian random variables with unit variance and zero mean, \textit{i.e.} for each $l,l' \in \mathbb{Z}$ we have
\begin{subequations}
 \label{eq:initcond}
 \begin{align}
	\overline{A_l} &=  \overline{B_l} = 0,\\
	\overline{A_{l'}A_l} &=  \overline{ B_{l'}B_l} = \delta_{l,l'},	\\
	\overline{A_{l'}B_l} &= 0,
 \end{align}
\end{subequations}
where the overline denotes the average of the random variables. As a consequence, each site $l$ of the grid representing the empty waveguide has the average atom density $\overline{|\psi_l(t_0)|^2}=1/2$.

We are now considering the source part which represent a BEC with a high number $N$ of atoms such that it can be safely described by a coherent state $\ket{\psi_S^0}$. The initial Wigner function $\mathcal{W}_S(\psi_S,\psi_S^*,t_0)$ therefore reads
\begin{equation}
\mathcal{W}_S(\psi_S,\psi_S^*,t_0) = \left(\frac{2}{\pi}\right) \exp(-2|\psi_S - \psi_S^0|^2).
\end{equation}
As $N$ is very large, the relative uncertainty of both the amplitude $|\psi_S^0| = \sqrt{N}$ and the associated phase of the source are negligibly small. We therefore treat the source term completely classically, \textit{i.e.} we set $\psi_S^0~=~\sqrt{N}$. 

Supposing, in addition, that the  coupling $\kappa(t)$ tends to zero such that $N|\kappa(t)|^2$ remains finite, we can safely neglect the depletion of the source or any back-action of the waveguide to the source since $\psi_S(t) = \sqrt{N}(1+\mathcal{O}(|\kappa|^2)$ at any finite time $t>t_0$. This allows us to solely focus on the evolution of the field in the chain. The equation to solve reads
\begin{eqnarray}
\label{eq:StochFinal}
i\hbar\frac{\partial \psi_l}{\partial t} &=& (V_l-\mu)\psi_l -J\left(\psi_{l+1} + \psi_{l-1}\right) \nonumber\\
					&&+ g_l(|\psi_l|^2 -1) \psi_l + \kappa(t)\sqrt{N}\delta_{l,l_S}.
\end{eqnarray}
One can notice that if $|\psi_l|^2$ is very large, we recover the discrete GP equation.

\subsection{Observables}
\label{subsec:Observables}
It can be shown \cite{Cahill1969PR} that the time-dependent expectation value of the symmetrically ordered product of the operator $\hat{a}_l$ and $\hat{a}^\dagger_l$ is of the form
\begin{eqnarray}
\label{eq:obscomputation}
\left\langle\left\{\prod_{l \in \mathbb{Z}} (\hat{a}_l^\dagger)^{r_l}\hat{a}_l^{s_l}\right\}_{\text{sym}}\right\rangle_t &=& \prod_{l \in \mathbb{Z}}\int d\psi_l d\psi^*_l\, (\psi_l^*)^{r_l}\psi_l^{s_l} \nonumber \\
	&& \times \mathcal{W}(\{\psi_l,\psi_l^*\},t),
\end{eqnarray}
where $\{ (\hat{a}_l^\dagger)^{r_l}\hat{a}_l^{s_l} \}_{\text{sym}}$ denotes the symmetrically ordered product \textit{i.e.} the average of $(r_l+s_l)!/(r_l!s_l!)$ possible orderings of $r_l$ creation operators and $s_l$ annihilation operators. For instance, setting $r_l=2$ and $s_l=1$ we have
\begin{equation}
  \left\{ (\hat{a}_l^\dagger)^2\hat{a}_l \right\}_{\text{sym}} = \frac{1}{3}\left[ (\hat{a}_l^\dagger)^2\hat{a}_l + \hat{a}_l^\dagger \hat{a}_l \hat{a}_l^\dagger + \hat{a}_l(\hat{a}_l^\dagger)^2 \right].
\end{equation}

This equation allows us to calculate the expectation value of observables on a particular site. Specifically, the expectation value of the total density $n_l(t)$ and the total current $j_l(t)$ on a site $l$ are given by
\begin{eqnarray}
n_l(t) &=& \langle\hat{n}_l(t)\rangle  = \langle\hat{a}_l^\dagger(t)\hat{a}_l(t)\rangle = \overline{|\psi_l(t)|^2} - 0.5, \\
j_l(t) &=& \langle\hat{j}_l(t)\rangle  \nonumber\\
	&=& \frac{i\hbar}{2}\left( \langle\hat{a}_{l+1}^\dagger(t)\hat{a}_l(t) - \hat{a}_l^\dagger(t)\hat{a}_{l+1}(t)\rangle\right) \nonumber \\
	  &=& \frac{i\hbar}{2}\left(   \overline{\psi^*_{l+1}(t)\psi_l(t) - \psi^*_l(t)\psi_{l+1}(t)}\right),
\end{eqnarray}
where the overline denotes the statistical average over all classical initial states. In addition, we can determine the coherent part of the density $n_l^{\text{coh}}(t)$ as well as the coherent part of the current $j_l^{\text{coh}}(t)$ through
\begin{eqnarray}
  n_l^{\text{coh}}(t) &=& |\langle\hat{a}_l(t)\rangle|^2 = \left|\overline{\psi_l(t)}\right|^2,\\
  j_l^{\text{coh}}(t) &=& \frac{i\hbar}{2}\left(   \overline{\psi^*_{l+1}(t)}\,\,\overline{\psi_l(t)} - \overline{\psi^*_l(t)}\,\,\overline{\psi_{l+1}(t)}\right).
\end{eqnarray}
In the mean--field limit, the coherent part of the density and the current correspond to the usual GP density and current. We can also identify the incoherent part as the difference of the total and coherent parts of the density and the current:
\begin{eqnarray}
 n_l^{\text{incoh}}(t) &=& n_l(t) - n_l^{\text{coh}}(t),\\
 j_l^{\text{incoh}}(t) &=& j_l(t) - j_l^{\text{coh}}(t).
\end{eqnarray}

The situation evidently simplifies in the special case of a waveguide without any on-site potential or interaction between the atoms. In this case, the GP as well as the tW evolution equations reduce to the standard one-body Schr\"{o}dinger equation and hence the coherent and total densities in the waveguide are identical. The stationary density $n^{\varnothing}$  is given by
\begin{eqnarray}
\label{eq:freen}
n^{\varnothing} &=& \lim_{t\to\infty} |\overline{\psi(t)}|^2 = \lim_{t\to\infty} \overline{|\psi(t)|^2} - 0.5 \nonumber \\
				&=& \frac{N|\kappa|^2}{4J^2-\mu^2}
\end{eqnarray}
and the stationary current $j^{\varnothing}$ is given by
\begin{equation}
\label{eq:freej}
j^{\varnothing} = \frac{N|\kappa|^2}{\sqrt{2J(4J^2-\mu^2)}}.
\end{equation}

\subsection{Truncated Wigner for open systems}
\label{subsec:tWSECS}
We are able to represent the infinite chain in terms of a finite open system if we assume that the on-site potential and the contact interaction are non-vanishing only in a finite region of space. This finite region will be named the \emph{scattering region} and the regions on the left and the right hand side of it are called the \emph{left} and \emph{right leads}, in close analogy to electronic mesoscopic physics. Without loss of generality, we shall assume that the scattering region is defined in the interval $l\in\{1,\cdots,L\}$ on the grid. The dynamics in the leads is linear and can therefore be solved analytically. We then find that the evolution equation can be written as \cite{Dujardin2014APB}
\begin{eqnarray}
\label{eq:integro_evol}
i\hbar \frac{\partial \psi_l}{\partial t} &=& (V_l-\mu)\psi_l + g_l |\psi_l|^2\psi_l + \kappa(t)\sqrt{N}\delta_{l,l_S} \nonumber\\
	& & - J\left[\psi_{l-1}(1-\delta_{l,1})+ \psi_{l+1}(1-\delta_{l,L})\right] \nonumber\\
	& & - \frac{2i}{\hbar}(\delta_{l,1} + \delta_{l,L})J^2 \int_{t_0}^t dt'\, \mathcal{M}_{1}(t-t')\psi_l(t') \nonumber\\
	& & +\delta_{l,1} \chi_1(t) + \delta_{l,L} \chi_L(t),
\end{eqnarray}
for the site $l$ within the scattering region ($l=1,\cdots,L$) with
\begin{subequations}
  \label{eq:classicalnoise}
  \begin{align}
    \chi_1(t) &= 2J \sum_{l'=-\infty}^0 \mathcal{M}_{l'-1}(t-t_0)\psi_{l'}(t_0), \\
	  \chi_L(t) &= -2J\sum_{l'=L+1}^\infty \mathcal{M}_{l'-L}(t-t_0)\psi_{l'}(t_0),
  \end{align}
\end{subequations}
and
\begin{equation}
\mathcal{M}_l(\tau) = \frac{i^l}{2} \left[ J_{l-1}
\left(\frac{2J\tau}{\hbar}\right)+J_{l+1}\left(\frac{2J\tau}{\hbar}\right)\right]e^{i\mu\tau/\hbar}
\end{equation}
where $J_l$ is the Bessel functions of the first kind of the order $l$.

As no approximation has yet been made, Eq.~\eqref{eq:integro_evol} reproduces the true evolution of the infinite nonlinear system under consideration described by Eq.~\eqref{eq:StochFinal}. The integral term in the third line of Eq.~\eqref{eq:integro_evol} exactly describes the decay into the left and right leads and therefore yields a perfectly transparent boundary condition that is defined on the first and last site of the scattering region. The terms $\chi_1(t)$ and $\chi_L(t)$ in Eq.~\eqref{eq:integro_evol} account for the propagation of the initial quantum fluctuations that arise in the framework of the tW approximtion and that eventually, during the time propagation, enter in the scattering region. These terms $\chi_1(t)$ and $\chi_L(t)$, considering the initial emptiness of the leads in the tW prescription (see Eqs.~\eqref{eq:InitCond}), take the form of \emph{quantum noise} entering the system. The autocorrelation functions related to these noise terms read are given by
\begin{equation}
  \overline{\chi_1^*(t)\chi_1(t+\tau)} = \overline{ \chi_L^*(t)\chi_L(t+\tau)} = -i \mathcal{M}_1(\tau).
\end{equation}

The integral term in Eq.~\eqref{eq:integro_evol} renders the numerical simulation rapidly inefficient because the whole integral has to be recomputed at every time step. The most efficient way to avoid this problem in the numerical computations \cite{Dujardin2014APB} is to remove this integral term and replace it by \emph{Smooth Exterior Complex Scaling} \cite{Balslev1971CMP,Simon1973AM,Simon1979PLA,Junker1982AAMP,Reinhardt1982ARPC,Ho1983PR,Loewdin1988AQC,Moiseyev1998PR}. The evolution of the finite open system is now governed by the following equation
\begin{eqnarray}
\label{eq:SECS}
i\hbar\frac{\partial \psi_l}{\partial t} &=& \left(V_l- \mu q_l\right)\psi_l + g_l(|\psi_l|^2-1) \psi_l \nonumber \\
    & &  + \kappa(t)\sqrt{N}\delta_{l,l_S} +2J( q_l + q_l^{-1}) \psi_l \nonumber \\
	& & -J\left[\frac{1}{q_{l+1}} +\frac{1}{2}\frac{q'_{l+1}}{ q^2_{l+1}}\right] \psi_{l+1} \nonumber \\
	& & -J \left[\frac{1}{q_{l-1}} -\frac{1}{2}\frac{q'_{l-1}}{ q^2_{l-1}}\right]\psi_{l-1} \nonumber \\
	& & +\delta_{l,1}\chi_1(t) + \delta_{l,L}\chi_L(t),
\end{eqnarray}
where $q_l$ is a smooth function of the site index $l$. In the scattering region ($1\leqslant l\leqslant L$) we impose $q_l=1$, while $q_l$ is smoothly ramped to $e^{i\theta}$ within the left ($l<1$) and the right ($l>L$) leads where $\theta$ is an arbitrary positive angle. The function $q'_l$ represents the discrete derivative of $q_l$ with respect to $l$. If $q_l\neq 1$, the Hamiltonian is not hermitian any longer and the outgoing atoms are absorbed without reflection, provided that the discrete function $q_l$ is sufficiently smooth (\textit{i.e.} $q_{l+1}-q_l\simeq q'_l\ll q_l$). This approach was successfully tested in Ref.~\cite{Dujardin2014APB} for the case of a linear and a nonlinear Schr\"odinger equation with or without quantum fluctuations as described in Eqs.~\eqref{eq:initcond} and Eqs.~\eqref{eq:classicalnoise}.

\section{Transmission across a quantum dot}
\label{sec:TransQD}
\subsection{Transmission spectrum}
\label{subsec:transmQD}
\begin{figure}[hbtp]
  \includegraphics[width=\linewidth]{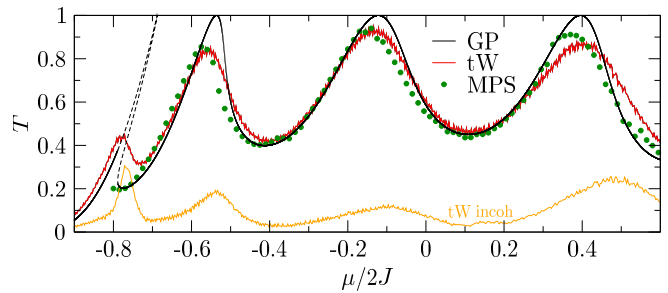} 
  \caption{(color online) Transmission across the quantum dot configuration versus $\mu/2J$ for $N|\kappa|^2 = J^2$, $g=0.2J$, $V=J$. The black curve corresponds to the mean--field (GP) calculation, the red curve to tW method and the green dots to the MPS method. We can see that for the GP method bistability occurs for the first resonance. The dashed black line depicts states that are not accessible during a time-dependent loading of the waveguide. The tW and MPS curves are in good agreement and exhibit an imperfect transmission at resonances. The incoherent part of the transmission is represented by an orange line. This curve shows that an appreciable amount of incoherent atoms are generated at the resonances, demonstrating a departure from of the GP model. The MPS method becomes numerically inefficient near the band edges, \textit{i.e.} for $\mu\approx -2J$.}
  \label{fig:V0.5g0.1}
\end{figure}

We now study transport across a symmetric double barrier potential that can be seen as a resonator. Hence, in absence of interaction, we know that the transmission spectrum will give rise to a series of Fabry-P\'{e}rot or Breit-Wigner peaks at resonances. As explained in Ref.~\cite{Paul2005PRL}, the presence of atom-atom contact interaction bends these peaks. Depending on the strength of the nonlinearity within the resonator, bistability can occur as seen in Fig.~\ref{fig:V0.5g0.1}. This bistability can be seen as an artifact of the mean--field approximation since many-body quantum scattering processes are linear from a microscopic point of view and, as a consequence, we expect a unique many-body scattering state to establish.

We now discuss the effects of the interaction on the Fabry-P\'{e}rot peaks beyond the mean--field GP description. We fix the maximal coupling strength between the source and the waveguide to $N|\kappa|^2 = J^2$. In Fig.~\ref{fig:V0.5g0.1}, we plot the transmission across the quantum dot against the chemical potential with an interaction strength $g=0.2J$ and an on-site potential $V=J$. The total transmission $T$ is determined by comparing the total current in the downstream region to the stationary current \eqref{eq:freej} obtained in the case of a perfectly homogeneous and interaction-free waveguide:
\begin{equation}
  \label{eq:totaltransmission}
  T = \lim_{t\to\infty} j(t)/j^{\varnothing}.
\end{equation}
It can be decomposed into its coherent $T^{\text{coh}}$ and incoherent $T^{\text{incoh}}$ part by respectively comparing the coherent and incoherent current to the free current \eqref{eq:freej}:
\begin{subequations}
 \begin{align}
	T^{\text{coh}} &= \lim_{t\to\infty} j^{\text{coh}}(t)/j^{\varnothing}, \\
	T^{\text{incoh}} &= \lim_{t\to\infty} j^{\text{incoh}}(t)/j^{\varnothing}.
 \end{align}
\end{subequations}
In the mean--field description, we observe that the GP curve is bent and features bistability as it was shown by Paul \textit{et. al.} \cite{Paul2005PRL}. This curve has been obtained by solving the stationary GP equation in the same way as it was done in Ref.~\cite{Paul2007PRA}. The dashed black curves correspond to solutions of the stationary GP equation that are unstable (middle branch of the resonance peak) or inaccessible through a time-dependent loading of the waveguide at constant chemical potential (upper branch of the resonance peak).

In order to benchmark our tW calculations, we compare the total transmission given by Eq.~\eqref{eq:totaltransmission} to the one obtained by a genuinely quantum simulation using Matrix-Product State (MPS) \cite{Vidal2003PRL,Verstraete2004PRL,Vidal2004PRL} calculations. This method is based on the Density-Matrix Renormalization Group \cite{White1992PRL} (DMRG) which uses renormalization techniques to express in an optimized way the density matrix of a block within the system under consideration. The states produced by this process belong to the class of matrix-product states \cite{Verstraete2004PRL,Vidal2004PRL,Vidal2003PRL}, which offer a highly optimized way of treating the full problem as long as no highly entangled states are present. When the number of atoms is quite low and the system is very small, the full Hilbert space can be efficiently truncated  by removing the degrees of freedom that are not involved in the dynamical evolution of the system. Such an optimized method enables us to numerically simulate the atomic quantum dot.

The results which are displayed in Fig.~\ref{fig:V0.5g0.1}, show a good agreement between the tW and MPS methods. Both methods clearly show that the transmission is not perfect at resonance, meaning that full resonant transmission is prohibited. The orange dotted curve in Fig.~\ref{fig:V0.5g0.1} displays the incoherent part of the transmission. We can see that about ten to twenty percent of the transmission comes from incoherent atoms at the resonances, which appears to be a consequence of the enhanced atomic density within the quantum dot at resonance. Indeed, in contrast to the coherent part of the transmitted beam, the incoherent atoms may exit the quantum dot to either one of the leads. They thereby inhibit perfect transmission of the atomic beam at resonance.

\subsection{Energy distribution of the transmitted atoms}
We are now interested in signatures of inelastic scattering in the transmitted beam. To this end, we take a large but finite number of sites $L_{\textrm{ft}}=1000$ in the transmitted region and define $\hat{a}(k)$ as
\begin{equation}
	\hat{a}(k) = \frac{1}{\sqrt{L_{\textrm{ft}}}}\sum_{l=L_D+2}^{L_D+2+L_{\textrm{ft}}} \hat{a}_l e^{-ikl},
\end{equation}
corresponding to the annihilation operator associated with the momentum eigenstate $e^{ikl}$ within the right lead. Noting that $[\hat{a}(k),\hat{a}^\dagger(k)]=1$ from this definition and following the procedure explained in the section \ref{subsec:Observables}, we can calculate the steady-state average total and coherent number of atoms moving with a wavenumber $k$ through
\begin{subequations}
 \begin{align}
 n(k) &= \langle\hat{n}(k) \rangle = \langle \hat{a}^\dagger(k)\hat{a}(k)\rangle=\overline{|\psi(k)|^2} - 0.5,\\
 n^\textrm{coh}(k) &=	|\langle \hat{a}(k) \rangle|^2= |\overline{\psi(k)}|^2,
 \end{align}
\end{subequations}
with
\begin{equation}
	\psi(k) = \frac{1}{\sqrt{L_{\textrm{ft}}}}\sum_{l=L_D+2}^{L_D+2+L_{\textrm{ft}}} \psi_l e^{-ikl}.
	\end{equation}
Since all the transmitted atoms have $k>0$, we define the total $n_E$ and coherent $n_E^\textrm{coh}$ average number of transmitted atoms moving with energy $E$ by
\begin{subequations}
 \begin{align}
 n_E &\equiv n(k_E),\\
 n_E^\textrm{coh} &\equiv n^\textrm{coh}(k_E),
 \end{align}
\end{subequations}
where $k_E$ is obtained by inverting the dispersion relation \eqref{eq:DispRel}:
\begin{equation}
  k_E = \arccos(-E/2J).
\end{equation}

\begin{figure}[hbtp]
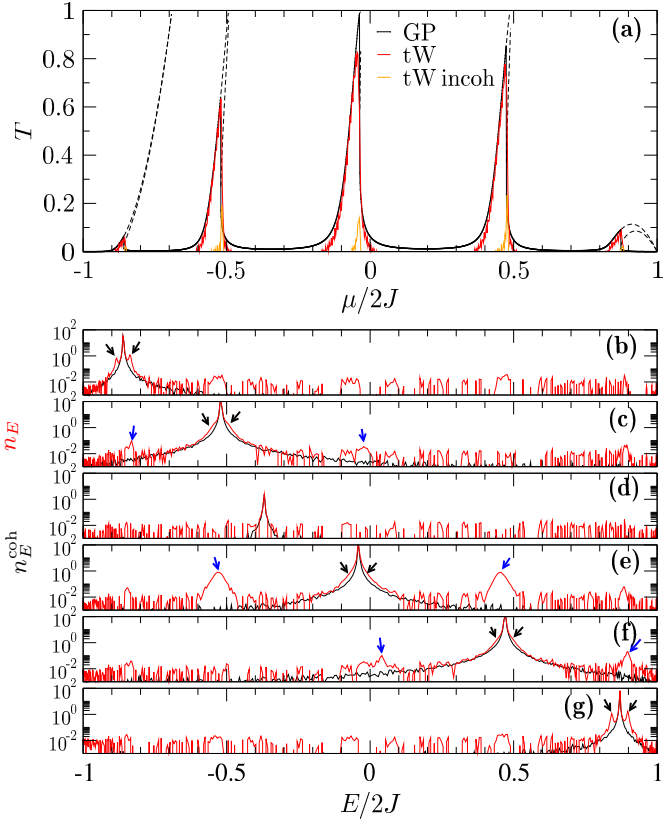

  \begin{overpic}[width=\linewidth]{eps/V2g0.01.eps}
    \put(-1,30){\rotatebox{90}{$n_E^{\textrm{coh}}$}}
    \put(-1,45){\rotatebox{90}{\textcolor{red}{$n_E$}}}
  \end{overpic}
  \caption{(color online) (a) Transmission spectrum of the quantum dot geometry with $V=4J$, $g=0.05J$, $N|\kappa|^2 = J^2$ computed by the GP (black line, following the same color convention as in Fig.~\ref{fig:V0.5g0.1}) and the tW prescription (red line for the total transmission and orange line for the incoherent part of the transmission). The panels below show the energy distribution of the outgoing flux according to the tW calculation for an incident beam energy of $\mu/2J=-0.86$ for (b), $\mu/2J=-0.52$ for (c), $\mu/2J=-0.37$ for (d), $\mu/2J=-0.04$ for (e), $\mu/2J=0.47$ for (f), $\mu/2J=0.87$ for (g). The black lines corresponds to the coherent and the red lines to the total part of outgoing atoms. The peaks designated by a black arrow arise from collective oscillations about a single resonance as discussed in Sec.~\ref{subsec:collosc}. The peaks designated by blue arrows arise from atoms that have undergone a transition between two single-particle levels of the atomic quantum dot.}
  \label{fig:V2g0.01}
\end{figure}

In Fig.~\ref{fig:V2g0.01}(a) we plot the transmission versus the normalized chemical potential of the incoming atoms with $V=4J$, $g=0.02J$ and $N|\kappa|^2 = \,J^2$. We can the see the appearance of well-resolved resonance peaks. Compared to Fig.~\ref{fig:V0.5g0.1} the visibility of the peaks is enhanced, which is expected as the enhancement of the potential barrier forming the quantum dot leads to a greater lifetime of the corresponding quasi-bound states. In Figs.~\ref{fig:V2g0.01}(b--g), we plot the energy distribution of the transmitted atoms. We can see the appearance of additional peaks depending on the value of the chemical potential. For Fig.~\ref{fig:V2g0.01}(d), where the chemical potential $\mu/2J=-0.37$ is far away from any resonance, we can only observe one peak corresponding to the coherent beam atoms coming from the source. In Fig.~\ref{fig:V2g0.01}(b,c,e--g) we can identify the appearance of two types of peaks (designated by arrows of different colors). As first type, we have two side peaks on the left- and right-hand side of the main peak at the incident beam energy, as seen for example in Fig.~\ref{fig:V2g0.01}(b,g) (black arrows). This will be further discussed in Sec.~\ref{subsec:collosc}. The second type of peaks correspond to inelastic scattering processes of atoms that thereby undergo a transition between different single-particle levels within the atomic quantum dot. They can be seen in Fig.~\ref{fig:V2g0.01}(c--f) (blue arrows) and will be discussed in Sec.~\ref{subsec:bogolqd}.

\subsection{Collective oscillations}
\label{subsec:collosc}
To understand the appearance of the two side peaks in the immediate vicinity of the incident beam energy, we study a leaky and driven single-level model with energy $E_0$ and two-body interaction between atoms with an interaction strength $g$. In the Heisenberg picture, the evolution equation of the field operator $\hat{a}\equiv\hat{a}(t)$ related to the level reads
\begin{equation}
\label{eq:SingLevQuanEvolEq}
	i\hbar \frac{\partial \hat{a}}{\partial t} = ( E_0 - i\gamma/2)\hat{a} + g \hat{a}^\dagger\hat{a}\hat{a} + [\kappa \hat{b} + \hat{\xi}(t)]e^{-i\mu t/\hbar},
\end{equation} 
where $\hat{b}$ correspond to the annihilation operator of the source which is coupled to the level with a coupling strength $\kappa$. The use of an imaginary leaky term $i\gamma/2$ implies that the losses are Markovian, which is justified in the limit of weak coupling between the single-level system and the leads. As was discussed in Sec.~\ref{subsec:tWSECS}, reducing the infinite waveguide to a finite open system introduces additional noise terms emerging from the initial vacuum fluctuations outside the quantum dot. Theses noise terms are accounted for by a time-dependent noise operator $\hat{\xi}(t)$ satisfying
\begin{equation}
  [\hat{\xi}(t), \hat{\xi}^\dagger(t)] = \xi_0^2\, \delta(t-t'),
\end{equation}
for some $\xi_0\in\mathbb{R}$. For the sake of simplicity we consider here a white noise. The commutation relations for the bosonic field operators are given by
\begin{subequations}
 \begin{align}
	[\hat{a}(0), \hat{a}^\dagger(0)] &= 1, \\
	[\hat{b}, \hat{b}^\dagger] &= 1.
 \end{align}
\end{subequations}
This model has the same ingredients as the atomic quantum dot system but offers the advantages to allow for analytical results. 

For this particular system, we are interested in the appearance of side peaks near the resonance for a weak atom-atom interaction and large population of the single-particle level. As a consequence, the truncated Wigner evolution equation of the wavefunction $\psi\equiv\psi(t)$ can be written as
\begin{equation}
\label{eq:SingLevClassEvolEq}
	i\hbar \frac{\partial \psi}{\partial t} = ( E_0 + g|\psi|^2 - i\gamma/2)\psi + [\kappa\sqrt{N}+\xi(t)]e^{-i\mu t/\hbar},
\end{equation}
where $N$ is the number of atoms in the source and the term $|\psi|^2-1$ is well approximated by $|\psi|^2$. The classical equivalent $\xi(t)$ of the quantum noise $\hat{\xi}(t)$ has following properties
\begin{subequations}
 \begin{align}
	\overline{\xi(t)} &= 0, \\
	\overline{\xi^*(t)\xi(t')} &= \frac{\xi_0^2}{2}\, \delta(t-t'),
 \end{align}
\end{subequations}
in perfect analogy with the truncated Wigner prescription to sample the initial quantum state with classical fields.

Instead of determining the number of atoms at energy $E$ by means of a spatial Fourier transform in the transmitted beam, we define it through a temporal Laplace transform of the amplitude on the level under consideration. We define the Laplace transform as 
\begin{equation}
	\label{eq:LaplaceTransform}
	\tilde{\psi}(E) = \frac{1}{\sqrt{\hbar T}}\int_0^\infty \, \psi(t) \exp\left[ \left( \frac{1}{T} - i \frac{E}{\hbar} \right) t \right]  dt,
\end{equation}
for a fixed (and ideally very large) observation time $T$. The number of atoms at energy $E$ is calculated according to Eq.~\eqref{eq:obscomputation} and reads
\begin{equation}
	\label{eq:tWObsPresc}
	\langle n_E\rangle = \overline{|\tilde{\psi}(E)|^2} - \frac{1}{2}[\tilde{a}(E),\tilde{a}^\dagger(E)].
\end{equation}

We are interested in collective oscillations of the condensate. For that purpose, we assume that we are close to a stationary state $\phi_0$ defined as the solution of the stationary GP equation
\begin{equation}
 (E_0-\mu- i\gamma/2 + g|\phi_0|^2) \phi_0 + \kappa\sqrt{N} = 0.
\end{equation}
We then decompose the wavefunction  $\psi(t)$ as $\psi(t)=(\phi_0 + \delta\psi(t))e^{-i\mu t / \hbar}$ and linearize the resulting evolution equation for $\delta\psi(t)$. We thereby obtain the Bogoliubov equations associated with Eq.~\eqref{eq:SingLevClassEvolEq} which read
\begin{equation}
 \label{eq:BogolClassicalSysEq}
 \left( 
  \begin{array}{cc}
   \Sigma-E & g\phi_0^2 \\
   -g\phi_0^{*^2} & -(\Sigma^*+E)
  \end{array}
 \right)
 \left( 
  \begin{array}{c}
   \delta\tilde{\psi}(E)  \\
   \delta\tilde{\psi}^*(-E)  
  \end{array}
 \right)
 =
 \left( 
  \begin{array}{c}
   -\tilde{\xi}(E)  \\
   \tilde{\xi}^*(-E)  
  \end{array}
 \right)
\end{equation} 
after applying a Laplace transform according to Eq.~\eqref{eq:LaplaceTransform}, with
\begin{equation}
	\Sigma=E_0-\mu+2g|\phi_0|^2 - i \left(\frac{\gamma}{2}-\frac{\hbar}{T}\right).
\end{equation}
Solving the system of equations \eqref{eq:BogolClassicalSysEq}, we find
\begin{equation}
	\overline{|\delta\tilde{\psi}(E)|^2} = \frac{(|\Sigma+E|^2 + g^2|\phi_0|^4)\xi_0^2 / 4\hbar}{|(\Sigma-E)(\Sigma^*+E) - g^2|\phi_0|^4|^2},
\end{equation} 
which yields 
\begin{equation}
	\overline{|\tilde{\psi}(E)|^2} = \frac{1}{\hbar T}\frac{|\phi_0|^2}{T^{-2} + (E-\mu)^2/\hbar^2} + \overline{|\delta\tilde{\psi}(E-\mu)|^2}.
\end{equation}
Following the same steps as in the previous lines, and supposing that $N\to\infty$, $\kappa\to0$ in such a way that $N|\kappa|^2$ remains constant, we can compute the commutator of Eq.~\eqref{eq:tWObsPresc}, which is given by
\begin{equation}
	[\tilde{a}(E),\tilde{a}^\dagger(E)] = \frac{(|\Sigma+E|^2 - g^2|\phi_0|^4)\xi_0^2 / 2\hbar}{|(\Sigma-E)(\Sigma^*+E) - g^2|\phi_0|^4|^2}.
\end{equation}
The total number of atoms at energy $E$ finally reads
\begin{eqnarray}
\label{eq:bogolne}
  \langle n_E \rangle &=& \frac{1}{\hbar T}\frac{|\phi_0|^2}{T^{-2} + (E-\mu)^2/\hbar^2} \nonumber \\
                      & & + \frac{g^2|\phi_0|^4\xi_0^2 / 2\hbar}{|(\Sigma-E)(\Sigma^*+E) - g^2|\phi_0|^4|^2}.
\end{eqnarray}

\begin{figure}[ht!]
  \begin{psfrags}
    \psfrag{theory}{theory}
    \psfrag{num. simulation}{num. simulation}
    \psfrag{0.1}{0.1}
    \psfrag{1}{1}
    \psfrag{-1}{-1}  
    \psfrag{0}{0}

    \psfrag{1e-03}[r][r]{$10^{\textrm{-}3}$}
    \psfrag{1e-02}[r][r]{$10^{\textrm{-}2}$}
    \psfrag{1e-01}[r][r]{$10^{\textrm{-}1}$}
    \psfrag{1e+00}[r][r]{$10^0$}
    \psfrag{1e+01}[r][r]{$10^1$}
    \psfrag{1e+02}[r][r]{$10^2$}
    \psfrag{1e+03}[r][r]{$10^3$}
    \psfrag{1e+04}[r][r]{$10^4$}
    \psfrag{1e+05}[r][r]{$10^5$}
    \psfrag{nE}{$\langle n_E\rangle$} 
    \psfrag{0.6}{0.6}
    \psfrag{0.4}{0.4}
    \psfrag{0.2}{0.2} 
    \psfrag{-0.2}{-0.2}
    \psfrag{-0.4}{0.4}
    \psfrag{-0.6}{-0.6}
    \psfrag{E-mu /E0}{$(E-\mu)/E_0$}    
    \psfrag{(a)}{\textbf{(a)}}          
    \psfrag{(b)}{\textbf{(b)}}          
    \psfrag{(c)}{\textbf{(c)}}              
    \includegraphics[width=\linewidth]{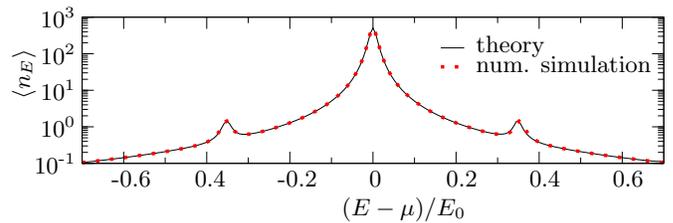}
  \end{psfrags}
  \caption{(color online) Average number of atoms $\langle n_E \rangle$ at energy $E$ for $g/E_0=0.02$, $\gamma/E_0=0.001$, $\sqrt{N}\kappa/E_0=0.05$, $E_0T=500\hbar$ and $\xi_0/E_0=0.5$. The presence of collective oscillations is clearly manifested in form of two side peaks appearing at $E-\mu\approx\pm0.36 E_0$. The (red) dots show the results obtained by numerically integrating Eq.~\eqref{eq:SingLevQuanEvolEq} and applying a Laplace transform according to Eq.~\eqref{eq:LaplaceTransform}. They are in perfect agreement with the theoretical prediction (black line) of Eq.~\eqref{eq:bogolne}.}
  \label{fig:BdG_1site}
\end{figure}

In Fig.~\ref{fig:BdG_1site}, we plot $\langle n_E \rangle$ for $\mu/E_0=1.08$ for an observation time $E_0T=500\hbar$ and $\xi_0/E_0=0.5$. The interaction strength is set to $g/E_0=0.02$, the leak rate to $\gamma/E_0=0.001$ and the source of atom to $\sqrt{N}\kappa/E_0=0.05$. We directly see the spectral signature of collective oscillations for $\mu/E_0=1.08$ which is close to the nonlinear resonance (\textit{i.e.} the population of the single-level system is high). This is in accordance with our previous findings for the quantum dot where collective oscillations appear near the resonances (see Fig.~\ref{fig:BdG_qdot}(a)). The occurrence of these side-peaks is, furthermore, in perfect qualitative agreement with the atom blockade study of Carusotto in Ref.~\cite{Carusotto2001PRA}.

\subsection{Bogoliubov excitations in the quantum dot}
\label{subsec:bogolqd}
We are now interested in the Bogoliubov modes within the multi-mode quantum dot configuration that we focus on in this paper. To this end, we numerically solve the Bogoliubov equations defined with respect to the stationary solution of the effective GP-like equation \eqref{eq:SECS}. The stationary wavefunction of Eq.~\eqref{eq:SECS} defined on the grid is given by $\phi_{0,l}$ on site $l$. We can solve the Bogoliubov equations
\begin{equation}
	\mathcal{T} \mathbf{y}^{(n)} = \epsilon_n \mathbf{y}^{(n)},
\end{equation}
where $\epsilon_n$ is the n$^\textrm{th}$ eigenvalue and $\mathbf{y}^{(n)}$ the related eigenvector. The matrix $\mathcal{T}$ is defined as
\begin{equation}
\mathcal{T} = 
 \begin{pmatrix}
	\mathcal{L} && \mathcal{C} \\
	-\mathcal{C}^* && -\mathcal{L}^*	
 \end{pmatrix},
\end{equation}
with the matrix elements of $\mathcal{L}$ and $\mathcal{C}$ defined by
\begin{eqnarray}
	\mathcal{L}_{ll'} &=& (V_l-\mu q_l+2g_l|\phi_{0,l}|^2)\delta_{l,l'} \\
	                  & & - J_{l'}(\delta_{l+1,l'} + \delta_{l-1,l'}) , \\
	\mathcal{C}_{ll'} &= &g\phi_{0,l}^2 \delta_{l,l'},
\end{eqnarray}
with $l,l^{'}= 0,1,\cdots, L$ and
\begin{equation}
  J_l =  J \left[\frac{1}{q_{l}} -\frac{1}{2}\frac{q'_{l}}{ q^2_{l}}\right],
\end{equation}
in the presence of SECS, see Eq.~\eqref{eq:SECS}. Clearly, $\mathcal{T}$ is not hermitian. Hence, the corresponding eigenvalues $\epsilon_n$ are complex, and their imaginary part is related to the width of the corresponding resonance peak.

\begin{figure}[ht!]
	\begin{center}
	\begin{overpic}[width=\linewidth]{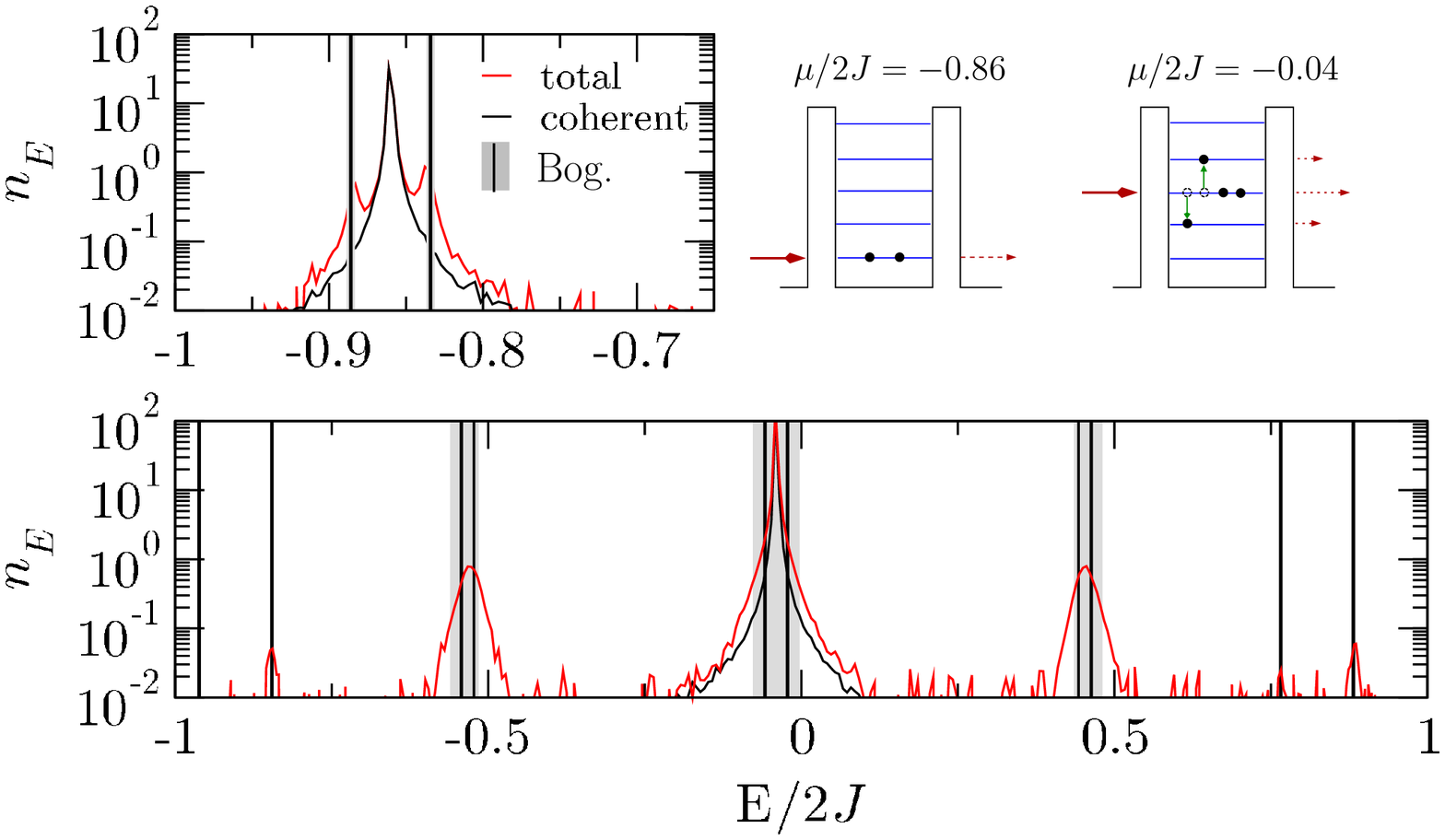}
	\end{overpic}
		\caption{(color online) Energy distribution of the transmitted beam for two different values of the chemical potential: $\mu/2J=-0.86$ for the upper panel and $\mu/2J=-0.04$ for the lower panel. The black vertical lines correspond to the expected Bogoliubov energies. They are in good agreement with the tW calculations. The grey zones correspond to the expected width of the peaks given by $2$Im($\epsilon_n$). For the upper panel, we see two side peaks around the chemical potential creating a collective oscillation inside the quantum dot. For the lower panel, we observe that, on top of the collective oscillation, inelastic collisions occur transferring atoms in the 2nd or 4th energy level as depicted in the sketch on the top right side. The tW results are well reproduced by the Bogoliubov theory.}
	\label{fig:BdG_qdot}
	\end{center}
\end{figure}

The numerically computed results are plotted for two different values of $\mu$ in Fig.~\ref{fig:BdG_qdot}. The vertical black lines correspond to the expected Bogoliubov eigenenergies Re$(\epsilon_n)$ and the grey zones correspond to the expected width of the peaks given by $2$Im($\epsilon_n$). The upper panel shows the results for $\mu/2J=-0.86$ and we can see that collective oscillations appear within the quantum dot in agreement with the Bogoliubov theory. The lower panel corresponds to a chemical potential $\mu/2J=-0.04$ that is close to the energy corresponding to the 3rd resonance. It shows a richer structure of peaks arising from the superposition of collective oscillations and inelastic scattering. Indeed, two colliding atoms at the incident energy $\mu/2J=-0.04$ can exchange energy through a collision process. After the collision, the first atom can end up on the 4th energy level and the second can end up on the 2nd energy level as depicted in Fig.~\ref{fig:BdG_qdot}. The results given in the tW calculation are in very good agreement with the Bogoliubov calculation. 

\section{Conclusions}
In the present work, we studied one-dimensional resonant transport of Bose--Einstein condensates within a guided atom laser configuration. For this purpose, we introduced a generalization of the truncated Wigner method to open systems. The reduction from an infinite system to a finite scattering region introduces an additional term accounting for quantum fluctuation which takes the form of a quantum noise. We made use of smooth exterior complex scaling to absorb the outgoing flux of atoms. This allowed us to study resonant and non-resonant transport across a one-dimensional atomic quantum dot beyond the mean--field Gross--Pitaevskii description.

The truncated Wigner method was used to compute the transmission across a quantum dot configuration. We observed that perfect resonant transmission is inhibited due to incoherent atoms creating a transmission blockade. This effect is in quantitative agreement with a Matrix-Product State calculation. The incoherent atoms originate from two different physical process. The first one is the creation of collective oscillations on an individual single-particle level within the quantum dot leading to two side peaks in the direct vicinity of the incident beam energy. The second one is related to inelastic collisions of atoms where atoms are transfered to other energy levels within the quantum dot.

The truncated Wigner method appears to be a very convenient tool to study transport of interacting Bose--Einstein condensates across more involved scattering configurations such as one--dimensional disordered potentials. This shall be discussed in a forthcoming publication \cite{Dujardin}. The approach presented in this paper can, furthermore, be extended to account for a more realistic description of the experimental configurations at hand involving, for instance, two reservoirs of $N$ atoms at ultralow but finite temperatures. This extension will then allow to  simulate source-drain transport processes across quantum dot like configurations, paving the way to a realistic theoretical study of atomtronics devices or atomic transistors.

\begin{acknowledgments}
The authors want to thank Boris Nowak for fruitful discussions. Computational resources have been provided by the Consortium des Equipements de Calcul Intensif (CECI), funded by the Fonds de la Recherche Scientifique de Belgique (F.R.S.-FNRS) under Grant No. 2.5020.11. 
\end{acknowledgments}

\bibliographystyle{bibtex/styles/prsty_all} 
\bibliography{bibtex/BEC,bibtex/AtomLaser,bibtex/ComScal,bibtex/Peter,bibtex/AndLoc,bibtex/TruncWig,bibtex/Transport,bibtex/CAP,bibtex/MPS,bibtex/Tronic,bibtex/Julien} 

\begin{thebibliography}{10}

\bibitem{Bloch1999PRL}
I. Bloch, T. H\"{a}nsch, and T. Esslinger, Phys. Rev. Lett. {\bf 82},  3008
  (1999).

\bibitem{Cennini2003PRL}
G. Cennini, G. Ritt, C. Geckeler, and M. Weitz, Phys. Rev. Lett. {\bf 91},
  240408  (2003).

\bibitem{Hagley1999S}
E.~W. Hagley, Science {\bf 283},  1706  (1999).

\bibitem{Mewes1997PRL}
M.-O. Mewes, M. Andrews, D. Kurn, D. Durfee, C. Townsend, {\it et~al.}, Phys.
  Rev. Lett. {\bf 78},  582  (1997).

\bibitem{Guerin2006PRL}
W. Guerin, J.-F. Riou, J. Gaebler, V. Josse, P. Bouyer, and A. Aspect, Phys.
  Rev. Lett. {\bf 97},  200402  (2006).

\bibitem{Couvert2008EEL}
A. Couvert, M. Jeppesen, T. Kawalec, G. Reinaudi, R. Mathevet, and D.
  Gu\'{e}ry-Odelin, EPL (Europhysics Letters) {\bf 83},  50001  (2008).

\bibitem{Riou2008PRA}
J.-F. Riou, Y. {Le Coq}, F. Impens, W. Guerin, C. Bord\'{e}, A. Aspect, and P.
  Bouyer, Phys. Rev. A {\bf 77},  033630  (2008).

\bibitem{Gattobigio2009PRA}
G. Gattobigio, A. Couvert, M. Jeppesen, R. Mathevet, and D. Gu\'{e}ry-Odelin,
  Phys. Rev. A {\bf 80},  041605  (2009).

\bibitem{Debs2010PRA}
J.~E. Debs, D. D\"{o}ring, P.~A. Altin, C. Figl, J. Dugu\'{e}, M. Jeppesen,
  J.~T. Schultz, N.~P. Robins, and J.~D. Close, Phys. Rev. A {\bf 81},  013618
  (2010).

\bibitem{KleineBuening2010APB}
G. {Kleine B\"{u}ning}, J. Will, W. Ertmer, C. Klempt, and J. Arlt, Appl. Phys.
  B {\bf 100},  117  (2010).

\bibitem{Gattobigio2011PRL}
G.~L. Gattobigio, A. Couvert, B. Georgeot, and D. Gu\'{e}ry-Odelin, Phys. Rev.
  Lett. {\bf 107},  254104  (2011).

\bibitem{Micheli2004PRL}
A. Micheli, A.~J. Daley, D. Jaksch, and P. Zoller, Phys. Rev. Lett. {\bf 93},
  140408  (2004).

\bibitem{Daley2005PRA}
A. Daley, S. Clark, D. Jaksch, and P. Zoller, Phys. Rev. A {\bf 72},  043618
  (2005).

\bibitem{Seaman2007PRA}
B.~T. Seaman, M. Kr\"{a}mer, D.~Z. Anderson, and M.~J. Holland, Phys. Rev. A
  {\bf 75},  023615  (2007).

\bibitem{Pepino2009PRL}
R.~A. Pepino, J. Cooper, D.~Z. Anderson, and M.~J. Holland, Phys. Rev. Lett.
  {\bf 103},  140405  (2009).

\bibitem{Brantut2012S}
J.-P. Brantut, J. Meineke, D. Stadler, S. Krinner, and T. Esslinger, Science
  {\bf 337},  1069  (2012).

\bibitem{Bruderer2012PRA}
M. Bruderer and W. Belzig, Phys. Rev. A {\bf 85},  013623  (2012).

\bibitem{Kristinsdottir2013PRL}
L. Kristinsd\'{o}ttir, O. Karlstr\"{o}m, J. Bjerlin, J. Cremon, P. Schlagheck,
  A. Wacker, and S. Reimann, Phys. Rev. Lett. {\bf 110},    (2013).

\bibitem{Carusotto2001PRA}
I. Carusotto, Phys. Rev. A {\bf 63},  023610  (2001).

\bibitem{Schlagheck2010NJoP}
P. Schlagheck, F. Malet, J.~C. Cremon, and S.~M. Reimann, New Journal of
  Physics {\bf 12},  065020  (2010).

\bibitem{Paul2005PRL}
T. Paul, K. Richter, and P. Schlagheck, Phys. Rev. Lett. {\bf 94},  020404
  (2005).

\bibitem{Paul2007PRA}
T. Paul, M. Hartung, K. Richter, and P. Schlagheck, Phys. Rev. A {\bf 76},
  063605  (2007).

\bibitem{Ernst2010PRA}
T. Ernst, T. Paul, and P. Schlagheck, Phys. Rev. A {\bf 81},  013631  (2010).

\bibitem{Geiger2012PRL}
T. Geiger, T. Wellens, and A. Buchleitner, Phys. Rev. Lett. {\bf 109},  030601
  (2012).

\bibitem{Geiger2013NJoP}
T. Geiger, A. Buchleitner, and T. Wellens, New Journal of Physics {\bf 15},
  115015  (2013).

\bibitem{Wigner1931}
E. Wigner, {\em Gruppentheorie und ihre Anwendung auf die Quantenmechanik der
  Atomspektren} (Vieweg+Teubner Verlag, Wiesbaden, 1931).

\bibitem{Wigner1932PR}
E.~P. Wigner, Phys. Rev. {\bf 40},  749  (1932).

\bibitem{Moyal1949PCPS}
J. Moyal, Proc. Cambridge Phil. Soc. {\bf 45},  99  (1949).

\bibitem{Scott2006PRA}
R. Scott, D. Hutchinson, and C. Gardiner, Phys. Rev. A {\bf 74},  053605
  (2006).

\bibitem{Scott2007LP}
R.~G. Scott, C.~W. Gardiner, and D.~A.~W. Hutchinson, Laser Physics {\bf 17},
  527–532  (2007).

\bibitem{Isella2006PRA}
L. Isella and J. Ruostekoski, Phys. Rev. A {\bf 74},  063625  (2006).

\bibitem{Altland2009PRA}
A. Altland, V. Gurarie, T. Kriecherbauer, and A. Polkovnikov, Phys. Rev. A {\bf
  79},  042703  (2009).

\bibitem{Schmidt2012NJoP}
M. Schmidt, S. Erne, B. Nowak, D. Sexty, and T. Gasenzer, New Journal of
  Physics {\bf 14},  075005  (2012).

\bibitem{Lee2014PRA}
M.~D. Lee and J. Ruostekoski, Phys. Rev. A {\bf 90},  023628  (2014).

\bibitem{Balslev1971CMP}
E. Balslev and J. Combes, Commun. Math. Phys. {\bf 22},  280  (1971).

\bibitem{Simon1973AM}
B. Simon, Ann. Math. {\bf 97},  247  (1973).

\bibitem{Simon1979PLA}
B. Simon, Phys. Lett. A {\bf 71},  211  (1979).

\bibitem{Kalita2011JCP}
D.~J. Kalita and A.~K. Gupta, J. Chem. Phys. {\bf 134},  094301  (2011).

\bibitem{Dujardin2014APB}
J. Dujardin, A. Saenz, and P. Schlagheck, Appl. Phys. B {\bf 117},  765–773
  (2014).

\bibitem{Glauber1963PR}
R.~J. Glauber, Phys. Rev. {\bf 131},  2766  (1963).

\bibitem{Sudarshan1963PRL}
E. Sudarshan, Phys. Rev. Lett. {\bf 10},  277  (1963).

\bibitem{Gardiner2004}
C.~W. Gardiner, {\em Handbook of stochastic methods for physics, chemistry and
  the natural sciences}, 3rd  ed. (Springer-Verlag, Berlin, 2004).

\bibitem{Olsen2009OC}
M. Olsen and A. Bradley, Opt. Commun. {\bf 282},  3924  (2009).

\bibitem{Sinatra2002JPBAMOP}
A. Sinatra, C. Lobo, and Y. Castin, J. Phys. B: At. Mol. Opt. Phys. {\bf 35},
  3599  (2002).

\bibitem{Cahill1969PR}
K.~E. Cahill and R. Glauber, Phys. Rev. {\bf 177},  1882  (1969).

\bibitem{Junker1982AAMP}
B. Junker, Adv. Atom. Mol. Phys. {\bf 18},  207  (1982).

\bibitem{Reinhardt1982ARPC}
W.~P. Reinhardt, Annu. Rev. Phys. Chem. {\bf 33},  223  (1982).

\bibitem{Ho1983PR}
Y. Ho, Phys. Rep. {\bf 99},  1  (1983).

\bibitem{Loewdin1988AQC}
P.-O. L\"{o}wdin, Adv. Quant. Chem. {\bf 19},  87  (1988).

\bibitem{Moiseyev1998PR}
N. Moiseyev, Phys. Rep. {\bf 302},  212  (1998).

\bibitem{Vidal2003PRL}
G. Vidal, Phys. Rev. Lett. {\bf 91},  147902  (2003).

\bibitem{Verstraete2004PRL}
F. Verstraete, D. Porras, and J. Cirac, Phys. Rev. Lett. {\bf 93},  227205
  (2004).

\bibitem{Vidal2004PRL}
G. Vidal, Phys. Rev. Lett. {\bf 93},  040502  (2004).

\bibitem{White1992PRL}
S.~R. White, Phys. Rev. Lett. {\bf 69},  2863  (1992).

\bibitem{Dujardin}
J. Dujardin, T. Engl, K. Richter, and P. Schlagheck, in preparation  .

\end{thebibliography}
\end{document}